\documentclass{aa}
\usepackage{psfig}
\usepackage{amssymb}

\advance\headerboxheight by -5mm

\def\l{l\kern -4.1pt\lower2pt\hbox{\tt \char19}\kern-0.8pt}
\def\L{L\kern -7pt\lower2pt\hbox{\tt \char19}\kern1pt}
\def\leer{~}
\begin{document}

\def\Msolar{{M$_{\odot}$\,}}
\def\arcsec{{$^{\prime\prime}$}\,}
\newcommand{\secpoint}{\mbox{$''\mskip-7.6mu.\,$}}
\newcommand{\minpoint}{\mbox{$'\mskip-5.6mu.\,$}}
\newcommand{\degpoint}{\mbox{$^\circ\mskip-7.6mu.\,$}}
\newcommand{\8}{S$_{\nu}^{8.64}$}
\newcommand{\4}{S$_{\nu}^{4.80}$}
\newcommand{\Ha}{H$\alpha$}
\newcommand{\Myr}{M$_{\odot}$~yr$^{-1}$}
\newcommand{\kms}{km~s$^{-1}$}
\newcommand{\vinf}{v$_{\infty}$}
\newcommand{\Zs}{Z$_{\odot}$}
\newcommand{\Md}{$\dot{M}$}
\newcommand{\Mi}{M$_{\circ}$}

\thesaurus{05(08.08.1; 08.12.3; 10.07.3 M~4)}

\title{
 The mass function of M~4 from near IR and optical HST
 observations\thanks{Based on observations with the
 NASA/ESA Hubble Space Telescope, obtained at the Space Telescope
 Science Institute, which is operated by AURA for NASA under contract
 NAS5-26555}}

\author{
 Luigi Pulone\thanks{On leave from  Osservatorio Astronomico di Roma, Via dell'Osservatorio 2, I-00040
 Monte Porzio Catone (Roma), Italy} \and Guido De{\leer}Marchi \and
 Francesco Paresce}
\authorrunning{L.~Pulone et al.}

\institute{
 European Southern Observatory, Karl-Schwarzschild-Strasse 2,
 D-85748 Garching, Germany\\
 E-mail: lpulone@eso.org, demarchi@eso.org, fparesce@eso.org}

\offprints{F.~Paresce}

\date{Received 2 October 1998 / Accepted 12 November 1998}
\maketitle

\begin{abstract}
 Deep images of the galactic globular cluster M\,4 taken at various
 locations with the NIC\,3 and the WFPC\,2 cameras on HST were used to
 derive detailed local optical and near IR luminosity functions.  White
 dwarfs have been detected for the first time on a color sequence at
 constant luminosity in the F110W band.  Transformation of the observed
 luminosity functions into mass functions via the most up to date
 theoretical mass luminosity relations currently available results in
 best fit local mass functions, in logarithmic mass units, that consist
 of a power-law $dN \propto m^{-x} d\log m$, with single exponent $x =
 -0.8$ for the inner regions and a two-segment power-law which rises
 with $x = 1$ down to $M_{814} \simeq 8.5$ and then drops all the way
 to the detection limit with $x = -0.3$ for the outer regions. This
 behavior cannot be reconciled with the expectations of a multi-mass
 King-Michie dynamical model using as input the canonical structure
 parameters for this cluster (core radius $r_{\rm c} = 50$\arcsec and
 concentration $c = 1.6$; Harris 1996). Thus, either the model does not
 accurately reflect the structure of the cluster due to some effect not
 properly accounted for in it or the canonical cluster structural
 parameters have to be significantly modified. Reasonable fits to all
 the present observations can be obtained with various global mass
 functions provided the cluster's structural parameters such as
 concentration $c$ and core radius $r_{\rm c}$ are in the range $c \in
 [1.4, 1.9]$ and $r_{\rm c} \in [50$\arcsec, $90$\arcsec]. The best
 compromise, in this case, consists in a model with a two--segment
 power--law mass function with exponents $x = 0.2\mbox{--}1.0 $ in the
 mass interval $0.8 < M_{\odot} < 0.25$, $x = -0.4$ for $m \leq 0.25
 M_\odot$ and structural parameters that require the least modification
 from the currently established values. This last result differs only
 minimally from that obtained for other globular clusters studied so
 far with HST which seem to have global mass functions increasing up to
 a peak at $\sim 0.25 M_\odot$ and then flattening out and possibly
 dropping to the H-burning limit.

\end{abstract}

 \section{Introduction}

 The galactic globular cluster M\,4 (NGC\,6121) is well suited to
 detailed measurements of its faint stellar population due to its
 proximity ($\sim 2$\,kpc) and looseness ($c = 1.59$; Harris 1996) that
 allows one, in principle, to probe its main sequence (MS) down to
 close to the H-burning limit with relative ease even from the ground
 (Kanatas et al. 1995). This capability is crucial in determining the
 shape of the stellar initial mass function (IMF) at the low mass end
 where observations have been notoriously sparse and unreliable
 (Chabrier \& M\`era 1997, De Marchi \& Paresce 1997). In particular, it
 is this part of the mass function (MF) close to the peak of the
 luminosity function that carries the most critical information
 concerning the possible log-normal form of the IMF (Adams \& Fatuzzo
 1996, Scalo 1998). Moreover, it is also here that one expects the
 effects of mass segregation and tidal stripping to be most severe and,
 therefore, most visible. If measurable, this signature could tell us
 much about the extent and nature of this phenomenon which is only
 imperfectly understood at the moment (Gnedin \& Ostriker 1997,
 Vesperini \& Heggie 1997).

 In practice, access to this region of the MS is particularly difficult
 due to a number of factors affecting the precision with which a
 luminosity function (LF) can be determined and a MF derived from the
 observed LF. Incompleteness, field star contamination, small number
 statistics, imperfect calibrations and, especially, very uncertain
 mass-luminosity relations (M-L) for low-mass stars have so far all
 conspired to severely limit the usefulness of M\,4 for this purpose.
 Kanatas et al. (1995) have come tantalizingly close to the end of the
 MS of M\,4 but the reliability of their results has been difficult to
 assess owing especially to the crudeness of the M-L relationship and
 to the inability to correct observationally for field star
 contamination. The most interesting aspect of their work has been the
 suggestion that this cluster is particularly poor in low mass stars
 which might imply that it has been significantly disrupted by internal
 and/or external forces. IR photometry from the ground carried out by
 Davidge \& Simons (1994) has also found a flat MF down to $0.2
 M_\odot$.

 In view of the recent availability of precise internally consistent
 M-L relationships for these metallicities (Baraffe et al. 1997) and
 the overwhelming importance to have at least one other cluster to
 compare with NGC\,6397, the only one for which, up to now, we had a
 reliable MF extending all the way down to the H-burning limit and to
 establish clearly the signature of tidal stripping on any cluster, we
 decided to observe M\,4 with the NICMOS instrument on board HST and to
 complement this extensive data set with archival WFPC\,2 observations
 in order to achieve a double goal: extend the LF reliably down to the
 H-burning limit and cover as wide an area in the cluster as possible
 to correctly account for the effects of mass segregation. In this
 paper, we present the results of this investigation. Details of the
 observations, their analysis and the color--magnitude diagrams are
 described in Sect.\,2 in which we compare the NICMOS observations with
 the WFPC\,2 results, showing how the infrared photometry has reached
 MS stars close to the H--burning limit, and we confirm the flattening
 of the MF for masses $ m < 0.25$\,\Msolar. The LFs and MFs are examined
 in Sect.\,3 whereas Sect.\,4 shows how the various LFs have been used
 to constrain a multi-mass King-Michie dynamical model of M\,4. A
 summary of the results follows in Sect.\,5.

 \section{Observations and analysis}

 The regions of the cluster analyzed in this paper are shown in
 Fig.\,1. The two HST-NIC3 images (dashed squares) were taken at $\sim
 1\minpoint9$ and $\sim 2\minpoint9$\,NE of the center. The two
 HST-WFPC\,2 archive fields used here were located $\sim 2^{\prime}$
 and $\sim 6^\prime$ E of the cluster center (``L'' shaped boxes). In
 the following, we describe these observations in detail.

 \begin{figure}
 \psfig{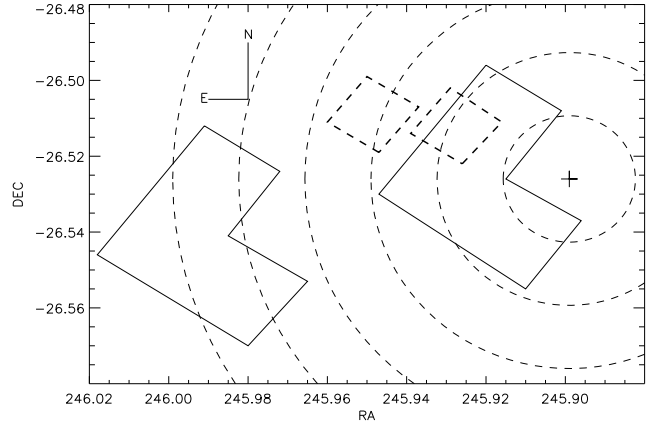}
 \caption[]{Location of all the fields used in this paper in M\,4 with
 respect to the cluster center marked with a cross.  The dashed radial
 circles start at 1$\arcmin$ with steps of 1$\arcmin$.  The dashed
 squares are the 51$^{\prime\prime}$ $\times$ 51$^{\prime\prime}$ wide
 NIC3 fields and the two fields bounded by the thin lines were observed
 through the WFPC2}
 \end{figure}

 \subsection{The NICMOS data}

 The observations of M\,4 described here were obtained on 1998
 January\,31 and February\,1 (UT) with the NICMOS instrument (MacKenty
 et al.  1997) on board the HST, during the first so called ``NIC\,3
 Campaign,'' when the HST secondary mirror was moved to place the
 NIC\,3 camera into focus. All the images were taken with this camera,
 which has a plate scale of $0\secpoint2$/pix, yielding a field of view
 of $51\secpoint2 \times 51\secpoint2$.

 Four exposures were taken at $RA = 16:23:42.6$ and $DEC = -26:30:40$,
 or $\sim 1\minpoint9$\,NE of the cluster center (field FN\,1). Four
 more exposures were taken, with the same instrumental setup, at $RA =
 16:23:47.6$, $DEC = -26:30:30$, or $\sim 2\minpoint9$\,NE of the
 cluster center (field FN\,2).  Finally, two additional images were
 taken at $RA = 16:24:04.7$, $DEC = -26:36:26$ (SKY field), i.e.
 $2^{o}$ away from the cluster center to account for the contamination
 due to field stars. All these images were obtained through the F110W
 and F160W filters using the MULTIACCUM mode. Table~1 summarizes the
 main characteristics of the images of these three fields.  For each
 observed field as indicated in column~1, we report the adopted filter,
 the exposure time in seconds, the MULTIACCUM predefined sample
 sequence, the number of exposures for each sample sequence and the
 number of MULTIACCUM sequences for each field.

 \begin{table}
 \caption[]{NIC 3 observations of the cluster M\,4}
 \begin{tabular}{@{}lcrccc@{}}
 \hline
 \noalign{\smallskip}
 $field$ &
 $filter$ &
 $t_{exp}$ &
 $Sampl.Seq.$ &
 $N.\,Sampl.$ &
 $N.\,exp$ \\
 \noalign{\smallskip}
 \hline
 \noalign{\smallskip}
 FN\,1 & F110W & 1919.87 & STEP 128 & 24 & $\times 2$ \\
 FN\,1 & F160W & 703.94 & STEP 28 & 20 & \\
 FN\,1 & F160W & 767.94 & STEP 28 & 24 & \\
 FN\,2 & F110W & 1919.87 & STEP 128 & 24 & $\times 2$ \\
 FN\,2 & F160W & 767.94 & STEP 28 & 24 & $\times 2$ \\
 SKY & F110W & 1919.87 & STEP 128 & 24 & $\times 2$ \\
 SKY & F160W & 703.94 & STEP 28 & 20 & \\
 SKY & F160W & 767.94 & STEP 28 & 24 & \\
 \noalign{\smallskip}
 \hline
 \end{tabular}
 \end{table}

 All frames have been processed with the standard NICMOS-HST pipeline
 procedure (CALNICA) to perform bias subtraction, linearity correction,
 dark count correction, flat fielding and cosmic ray identification and
 rejection. The frames in the same field and taken through the same
 filter have been registered with respect to each other and averaged to
 improve the statistics. The final images correspond to a total
 exposure time of 64 min for the F110W filter in all the observed
 fields, 24 min for the F160W filter in FN\,1 and SKY and 27 min for the
 F160W filter in FN\,2.

 The FN\,1, FN\,2 and SKY combined frames are shown in Figs.\,2, 3, and
 4, respectively. The image quality is excellent and the background is
 almost flat with the exception of the regions surrounding bright
 stars. The average per pixel count rate is $0.030$\,count $s^{-1}$ and
 $0.035$\,count $s^{-1}$, respectively in F110W and F160W, with
 standard deviation of $0.015$\,count $s^{-1}$ and $0.011$\,count
 $s^{-1}$.

 \begin{figure}
 \vbox{\hbox to\textwidth{\psfig{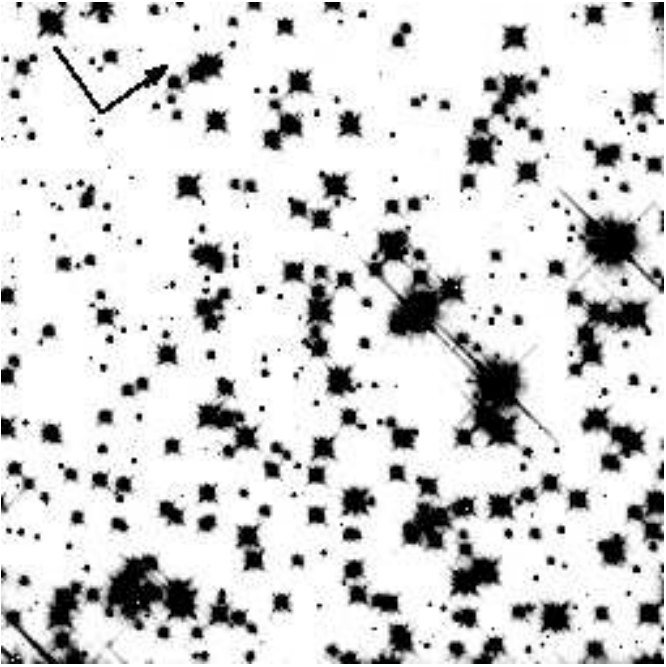}\hfill}}
 \caption[]{Negative image of the field FN\,1 in the globular cluster
 M\,4, located $\sim 1\minpoint9$ away from the nominal cluster
 center, as observed with the F110W filter of the NIC\,3 camera. The
 field covers $\sim 51$\arcsec on a side. North is indicated by the
 arrow in the upper-left corner with East indicated to the left of
 North. The exposure time of the combined image is 3,840\,s}
 \end{figure}

 \begin{figure}
 \vbox{\hbox to\textwidth{\psfig{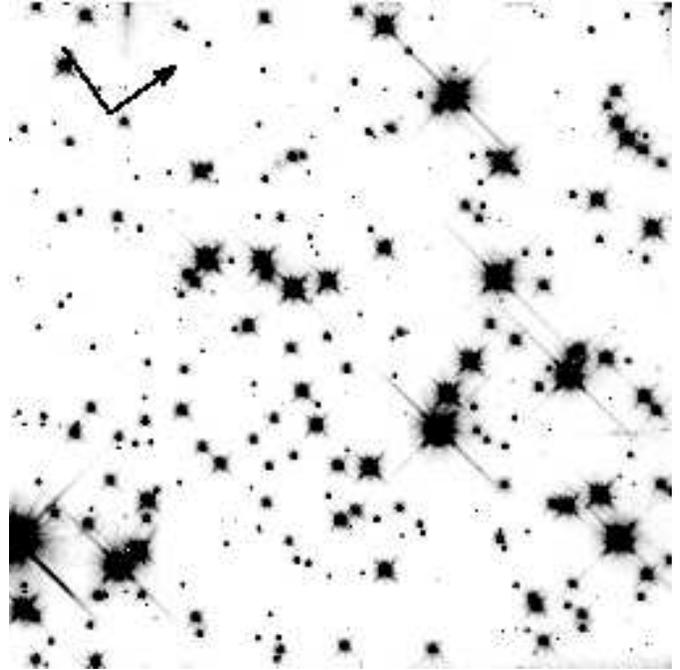}\hfill}}
 \caption[]{Same as Fig.\,2 but for Field FN\,2,
 located $2\minpoint9$ away from the center of the cluster}
 \end{figure}

 \begin{figure}
 \vbox{\hbox to\textwidth{\psfig{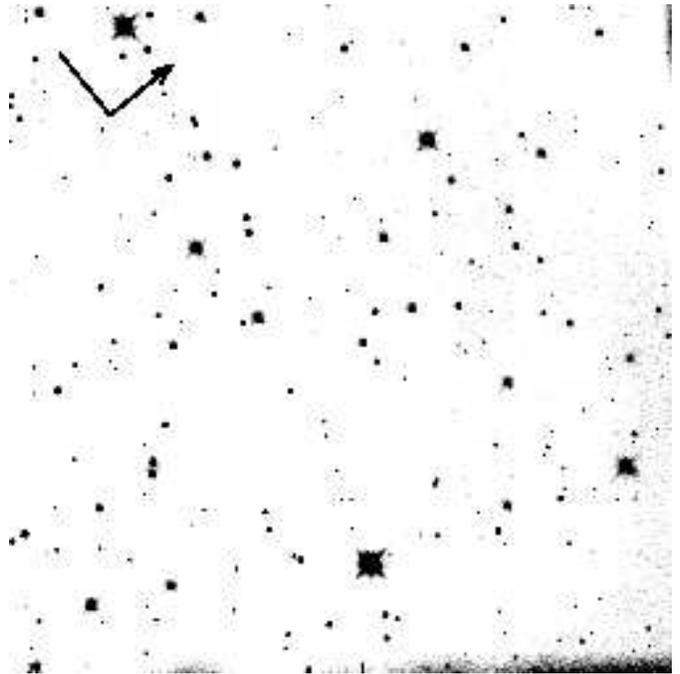}\hfill}}
 \caption[]{Same as Fig.\,2 but for Field SKY}
 \end{figure}

 Photometry was carried out using the standard IRAF DAOPHOT package.
 The automated star detection routine {\it daophot.daofind} was applied
 to the frames, by setting the detection threshold at $5\,\sigma$ above
 the local background level. We have then examined by eye each individual
 object detected by daofind and discarded a number of features (PSF
 tendrils, noise spikes, etc.) that daofind had misinterpreted as stars,
 as well as a few extended objects whose full width at half maximum
 (FWHM) exceeded by a factor of two or more that typical of stellar
 objects in our frames (FWHM $\simeq 1.3$\,pixel). As regards the
 saturation of bright objects, the CALNICA pipeline verifies whether and
 when the signal in each pixel, in the multiple readouts of each
 MULTIACCUM sequence, reaches the saturation or nonlinearity limit prior
 to the end of the exposure. All the samples which occur after the
 saturation limit are identified as bad and are not used in the final
 calculation of the mean count rate. This procedure should, in
 principle, allow one to measure fluxes in a wide dynamic range
 ($\sim\,10$ mag in our case).

 A total of $466$, $332$ and $175$ objects respectively in the FN\,1,
 FN\,2 and SKY fields were identified and their fluxes measured with the
 PSF-fitting {\it daophot.allstar} photometry package, to derive the
 color--magnitude diagrams (CMD) for the three fields. The internal
 photometric accuracy has been estimated both with the IRAF {\it phot}
 task and through standard artificial--star experiments, giving similar
 results: better than $\pm 0.02$ mag for bright stars, degrading to
 about $\pm0.3$ mag toward the detection limit at $ m_{110} \sim27$.

 Instrumental magnitudes were calibrated and converted into the HST
 magnitude system (STMAG) using the relation:
 \begin{equation}
 m_{ST}= -2.5 \log\left(\frac{c \,U}{\varepsilon}\right) -21.1
 \end{equation}
 where $c$ is the count rate measured for each star, $U$ the inverse
 sensitivity of the instrumental setup (camera + filters), and
 $\varepsilon$ the encircled energy (i.e. the fraction of the total
 flux sampled by our photometry). In order to estimate the encircled
 energy $\varepsilon$, we have used five isolated, relatively bright
 stars, and have compared the flux measured by our photometry to their
 total flux, defined as that falling within a $2$\arcsec radius. We
 have found in this way that, for our choice of aperture radius ($2.5$
 pixel) and background annulus ($2.5\mbox{--}10$ pixel) the encircled
 energy amounts to $\simeq 90\,\%$ for both the F110W and F160W
 filters. Because the calibration of the NIC\,3 camera is still
 preliminary, and the values of the inverse sensitivity $U$ are based
 on a limited set of spectrophotometric standards, we cannot convert
 very accurately our magnitudes from the STMAG system into any
 ground-based system. We, therefore, expect our absolute photometry to
 be accurate to within $10\mbox{--}15$\,\% (Colina \& Rieke 1997). It
 is still possible, however, to translate our measurements into the
 VEGAMAG photometric system of the HST, defined as one in which the
 magnitude of Vega would be 0 in all bands and which is more similar to
 the classical Johnson--Cousin ground-based system. This can be done by
 simply subtracting the zero-point constants of $2.3$\,mag and
 $3.7$\,mag from the values of $m_{110}$ and $m_{160}$, respectively,
 thus obtaining the $J$ and $H$ magnitudes that we use hereafter.

 The CMDs of all the objects identified in the three fields observed
 (FN\,1, FN\,2 and SKY) are shown in Figs.\,5, 6, and 7. The MS is well
 defined in both cluster fields down to $J \simeq 20$, but below this
 level the stellar density decreases abruptly. In the SKY field,
 located $2^o$ away from the cluster center, the density of stars
 begins to increase for $J > 20$, implying that the stellar population
 of the cluster becomes negligible at this point. Using the M--L
 relation of Baraffe et al. (1997), this point corresponds to $\sim
 0.2$\,\Msolar for a distance modulus of $(m-M)_{\rm J} = 11.86$.
 Another noticeable feature of the MS is the change of slope at $J
 \simeq 18$ corresponding to $M \simeq 0.5 M_\odot$.  This is because
 of the increasing infrared absorption of $H_2$ molecules in the
 stellar atmosphere which counteracts the reddening due to both the
 decreasing temperature and increasing metallic molecular absorption in
 the optical band (Baraffe et al. 1997). Below this limit, the MS
 becomes almost vertical.

 The broadening of the MS for $ J < 16.5$ is greater than expected on
 the basis of the photometric errors alone.  This broadening has been
 already noticed by Richer et al. (1995).  Although it cannot be
 excluded that this feature is partly due to differential reddening
 (Alcaino et al. 1997) or to the presence of a binary sequence (Kroupa
 \& Tout 1992; Ferraro et al. 1997), it seems more likely that this
 effect has to be ascribed to the non uniformity in the pixel response
 function of the camera (intra-pixel sensitivity) which has been shown
 to affect the photometry at the $10\%$ level in NIC\,3 (Storrs 1998).
 On the other hand, the CMD of M\,4 published by Kanatas et al. (1995)
 does not show this peculiar MS broadening. We have estimated that if
 the MS spread were to depend on the data processing alone, the actual
 error should be almost four times larger than the photometric
 uncertainty ($\sim 0.04$ mag) in the range $J < 16.5$. As a
 consequence, it would be difficult to evaluate the age of the cluster
 from these data, but the LF of MS stars is little affected by this
 systematic effect.

 We have superposed on the data shown in Figs.\,5 and 6 the expected
 10\,Gyr, [M/H]$=-1.0$ theoretical isochrone (Chabrier \& Baraffe 1997;
 Allard \& Hauschildt 1997), as seen through the NICMOS filters F110W
 and F160W. For this comparison we have adopted for M\,4 a color excess
 E(B--V)$=0.40$ and a distance modulus of $(m-M)_{\rm o} = 11.51$
 (Djorgovski \& Meylan 1993). The ensuing values of the absorption in
 the F110W and F160W bands, estimated with the relation of Cardelli,
 Clayton, \& Mathis (1989), are respectively $A_{110} = 0.33$ and
 $A_{160} = 0.18$.  Although these values do not bring the theoretical
 models into perfect agreement with the observed CMD (an additional
 $\sim 0.1$\ mag shift to the red seems to be required), we regard this
 mismatch as insignificant in light of the uncertainties on both the
 cluster's parameters and theory. Moreover, there are indications of
 higher and variable values for $R_{V}$ (Vrba, Coyne, \& Rapia 1993)
 towards the direction of M\,4 which is behind the Sco--Oph dust
 complex. Thus, because the calibration of our filters is still
 preliminary, the match between observation and theory as shown in both
 the FN\,1 and FN\,2 fields (Figs.\,5 and 6) is remarkable.

 \begin{figure}
 \vbox{\hbox to\textwidth{\psfig{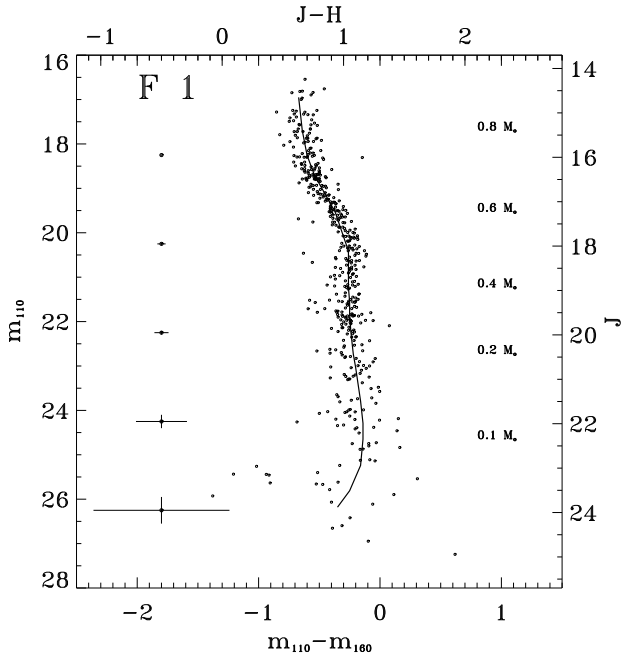}\hfill}}
 \caption[]{Color-magnitude diagram of the field FN\,1 located
 $1\minpoint9$ off the center of the globular cluster M\,4 (NGC\,6121).
 A total of 466 stars are detected and measured in two bandpasses
 (F110W and F160W) with the NIC\,3 camera in this field. The solid line
 represents the expected theoretical MS from Baraffe et al. (1997).
 The magnitudes and colors in the HST-VEGAMAG system are shown
 respectively on the right and on the upper axis of the diagram.
 Photometric error bars are indicated on the left.  The stellar masses
 determined by the M-L relationship of Baraffe et al. (1997) are
 indicated on the right side of the diagram}
 \end{figure}

 \begin{figure}
 \vbox{\hbox to\textwidth{\psfig{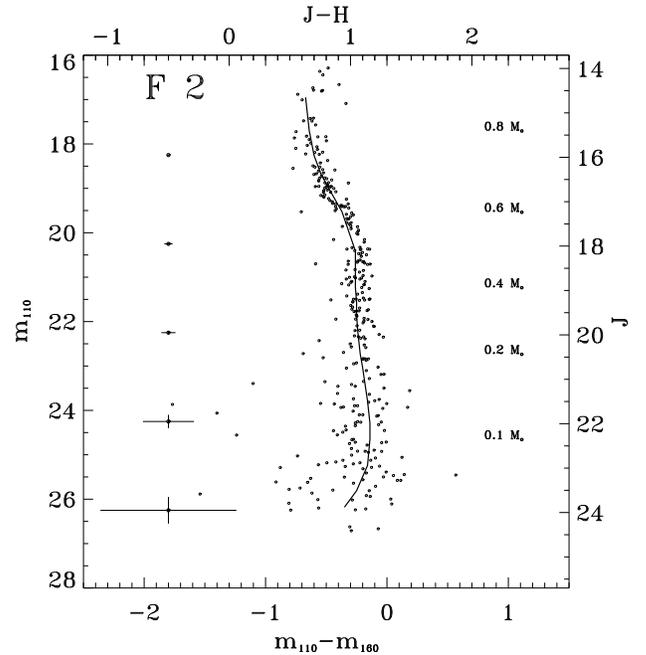}\hfill}}
 \caption[]{Same as Fig.\,3 but for field FN\,2, located $2\minpoint9$
 away from the center of the cluster.
 A total of 332 objects are measured in this field}
 \end{figure}

 \begin{figure}
 \vbox{\hbox to\textwidth{\psfig{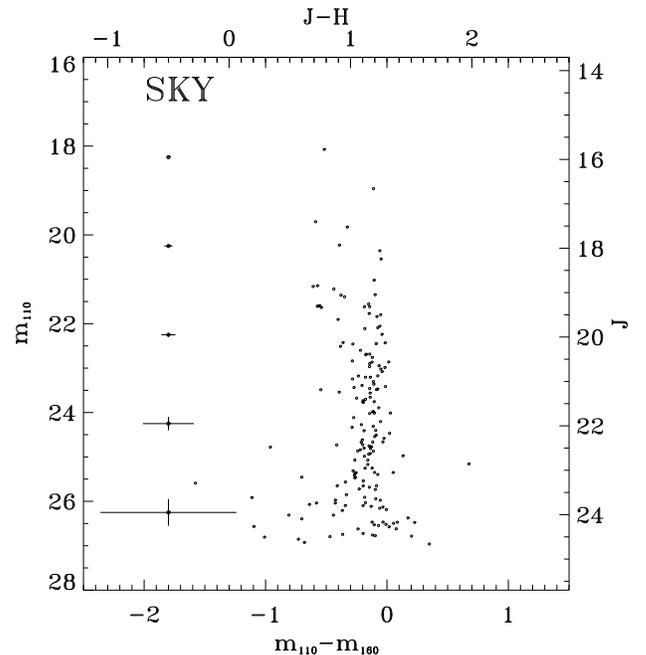}\hfill}}
 \caption[]{CMD of the 175 objects in the field SKY at $\sim 2^{o}$
 from the center of the cluster}
 \end{figure}

 The completeness of the sample has been evaluated by running
 artificial star tests in both bands. For each $0.5$ magnitude bin we
 have carried out 10 trials by adding a number of artificial stars not
 exceeding 10\,\% of the total number of objects in that bin and using
 a PSF derived directly from the co-added frames. These trials were
 followed recursively by daofind and allstar runs with the same
 parameters used in the reduction of the scientific images to assess
 the fraction of objects recovered by the procedure and the associated
 photometric error. Moreover, for each artificially added star, we
 compared the input magnitude with that at which the star was
 recovered, and found no systematic difference. In addition, the
 standard deviation of the magnitude values at which stars are
 recovered is in excellent agreement with the $1\,\sigma$ uncertainty
 of the photometric reduction, as estimated by the IRAF {\it phot}
 task, at all magnitude levels and is indicated on the left-hand side
 of the CMDs shown in Figs.\,5, 6, 7.

 The photometric completeness of the three observed fields is displayed
 in Fig.\,8 as a function of the $m_{110}$ magnitude, whereas Table~2
 lists, for each magnitude bin, the completeness fraction, the
 corrected LF, and the corresponding rms errors coming from the
 poissonian statistics of the counting process (all values of the LFs
 as well as their errors have been rounded off to the nearest
 integer).

 \begin{table}
 \caption[]{Luminosity function and completeness}
 \begin{tabular}{@{}rrclcrclcrclc@{}}
 \hline
 \noalign{\smallskip}
 & \multicolumn{3}{c}{FN\,1} &
 \multicolumn{3}{c}{FN\,2} &
 \multicolumn{3}{c}{SKY} \\
 $m_{110}$ &\% &
 $N$ &
 $\sigma_{N}$ &\% &
 $N$ &
 $\sigma_{N}$ &\% &
 $N$ &
 $\sigma_{N}$\\
 \noalign{\smallskip}
 \hline
 \noalign{\smallskip}
 17.25 & 100 & 14 & 4 & 100 & 6 & 2 & 100 & 0 & 0\\
 17.75 & 99 & 26 & 4 & 99 & 9 & 3 & 100 & 0 & 0\\
 18.25 & 98 & 30 & 6 & 99 & 11 & 3 & 100 & 1 & 1\\
 18.75 & 97 & 54 & 8 & 98 & 28 & 5 & 100 & 1 & 1\\
 19.25 & 95 & 40 & 7 & 98 & 26 & 5 & 99 & 0 & 0\\
 19.75 & 94 & 41 & 7 & 96 & 19 & 4 & 99 & 2 & 1\\
 20.25 & 92 & 35 & 6 & 96 & 18 & 4 & 98 & 2 & 1\\
 20.75 & 89 & 36 & 7 & 95 & 20 & 5 & 98 & 1 & 1\\
 21.25 & 87 & 26 & 6 & 94 & 23 & 5 & 98 & 7 & 3\\
 21.75 & 83 & 55 & 9 & 91 & 24 & 5 & 98 & 11 & 3\\
 22.25 & 82 & 29 & 6 & 90 & 16 & 4 & 97 & 8 & 3\\
 22.75 & 78 & 27 & 6 & 88 & 16 & 4 & 97 & 13 & 4\\
 23.25 & 74 & 22 & 6 & 85 & 15 & 4 & 96 & 17 & 4\\
 23.75 & 70 & 17 & 5 & 81 & 21 & 5 & 94 & 13 & 4\\
 24.25 & 66 & 26 & 7 & 75 & 23 & 6 & 93 & 12 & 4\\
 24.75 & 58 & 13 & 5 & 66 & 23 & 6 & 88 & 25 & 6\\
 25.25 & 51 & 18 & 7 & 66 & 33 & 8 & 84 & 19 & 5\\
 25.75 & 41 & 20 & 7 & 55 & 32 & 9 & 78 & 19 & 5\\
 26.25 & 32 & 9 & 6 & 37 & 22 & 9 & 66 & 30 & 8\\
 26.75 & 20 & 15 & 11 & 26 & 11 & 9 & 51 & 33 & 10\\
 \noalign{\smallskip}
 \hline
 \end{tabular}
 \end{table}

 Because we planned the NICMOS exposures in such a way that the F110W
 and F160W band images reach the same limiting magnitude with the same
 SNR, and because the colors of all the objects seen in the CMD (both
 cluster and field stars) span a very narrow range, all the objects
 identified in the F110W frames are also found in the F160W images
 and, therefore, the photometric completeness is similar in both bands.

 \begin{figure}
 \vbox{\hbox to\textwidth{\psfig{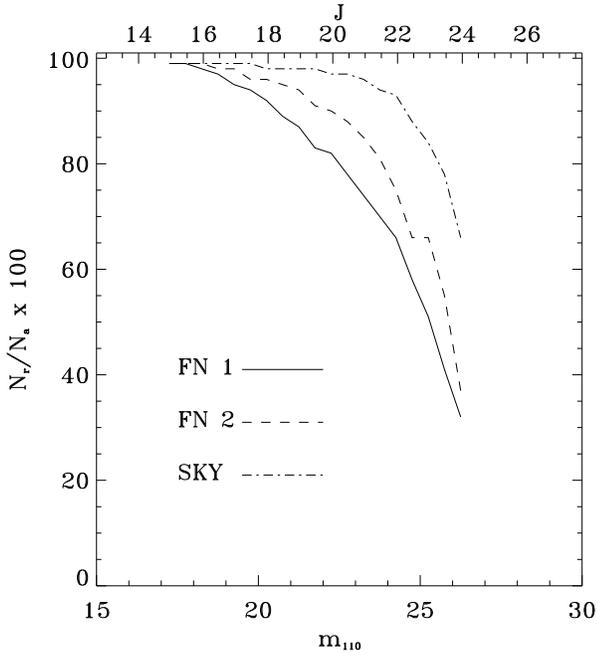}\hfill}}
 \caption[]{Photometric completeness for the three observed fields.
 The completeness was calculated as the ratio of the numbers of the
 recovered stars $N_r$ and the added artificial stars $N_a$}
 \end{figure}

 To determine the stellar LF of M\,4 at the locations corresponding to
 our fields FN\,1 and FN\,2 we have used the color and magnitude
 information derived from the CMDs in Figs.\,5 and 6. Because M\,4 is
 projected towards the Galactic bulge at a low Galactic latitude ($b
 \simeq 16$), the contamination due to field stars in FN\,1 and FN\,2
 is expected to be significant. While at optical wavelengths it is
 often possible to disentangle the stars belonging to the cluster from
 those that do not simply on the basis of their color (see next section
 for a description of the ``$2.5\,\sigma$ clipping'' criterion), the
 colors of low-mass stars in the near-IR make this a very difficult
 undertaking. And indeed, both cluster and field stars share the same
 narrow color region in our F110W, F160W CMDs (see Figs.\,5, and 6),
 with the latter scattered in luminosity because of their varying mass
 as well as distance. We had, however, expected this behavior (see
 Baraffe et al. 1997), and had therefore obtained a comparison field
 (SKY) at the same projected distance from the Galactic bulge as the
 FN\,1 and FN\,2 fields. Thus, the use of a control field is mandatory
 to reliably measure the stellar LF of this cluster, due to its
 unfavorable location.

 \begin{figure}
 \vbox{\hbox to\textwidth{\psfig{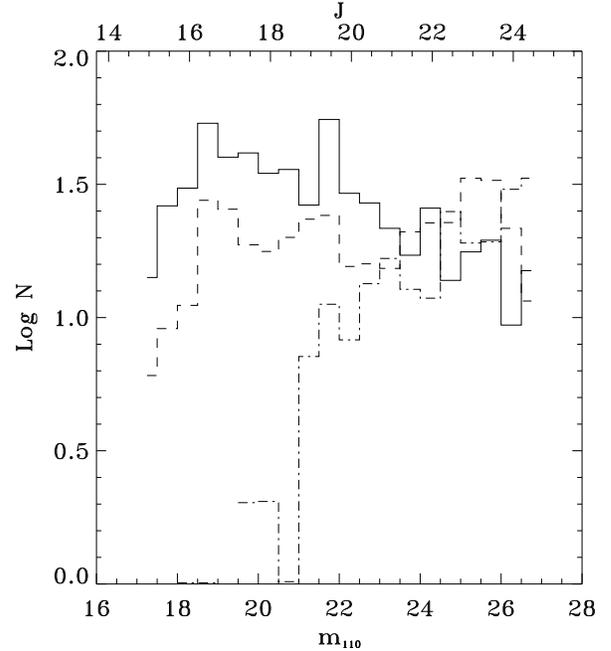}\hfill}}
 \caption[]{The luminosity functions corrected for photometric
 incompleteness in
 the three observed fields, FN\,1 (solid line), FN\,2 (dashed line)
 and SKY (dash-dotted line)}
\vspace{4mm}
 \end{figure}

 We have measured the number of stars in each $0.5$ magnitude bin as a
 function of the $m_{110}$ magnitude along the MS in all three CMDs,
 discarding stars bluer than $J - H \simeq 0.4$ as they are clearly not
 MS objects. We have adopted exactly the same color selection also for
 the SKY field, so as to ensure the consistency of the comparison with
 the FN\,1 and FN\,2 fields data.  Fig.\,9 shows the LFs in the three
 observed fields FN\,1 (continuous line), FN\,2 (dashed line) and SKY
 field (dot-dashed line) after correction for photometric
 incompleteness as reported in Fig.\,8 and Table~2. The artificial star
 tests indicate that at $J \sim 23$ on average more than 50\,\%,
 60\,\%, and 80\,\% of the added stars are recovered respectively in
 the FN\,1, FN\,2, and SKY, thus implying statistically reliable LFs
 towards the H-burning limit at $\sim 0.1 M_{\odot}$, corresponding to
 $ J = 22.5$ according to the M--L relations of Baraffe et al. (1997).
 In spite of the different crowding level and photometric completeness,
 the LFs measured in the adjacent fields FN\,1 and FN\,2 match each
 other very well, both showing peaks at $J \simeq 16.5$ and $J \simeq
 19.5$. It is interesting to note that, after correction for
 incompleteness, all three LFs tend to a plateau for $J > 21$, clearly
 marking in this way the bottom of the MS, from where field star
 contamination increases.

 \begin{figure}
 \vbox{\hbox to\textwidth{\psfig{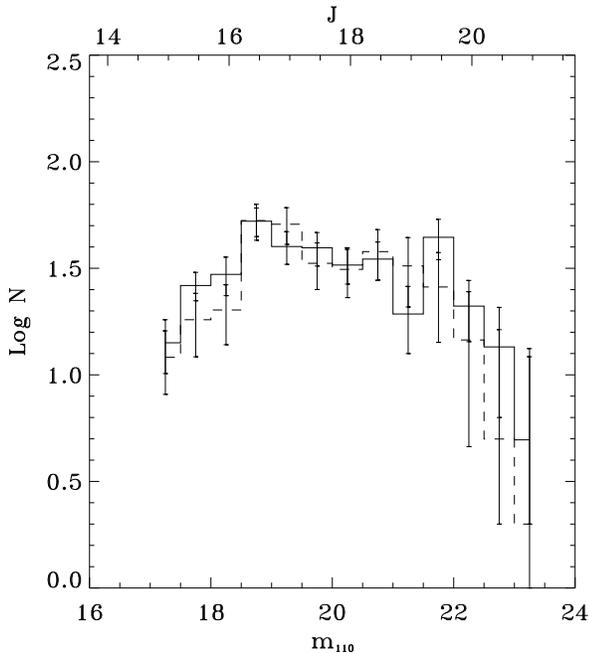}\hfill}}
 \caption[]{IR luminosity functions of M\,4, obtained from our
 photometry and corrected for both incompleteness and contamination.
 The solid histogram represents the inner field (FN\,1) LF
 whereas the dashed histogram reflects the outer field (FN\,2).}
 \end{figure}

 \begin{figure*}
\vbox{\hbox
to\textwidth{\psfig{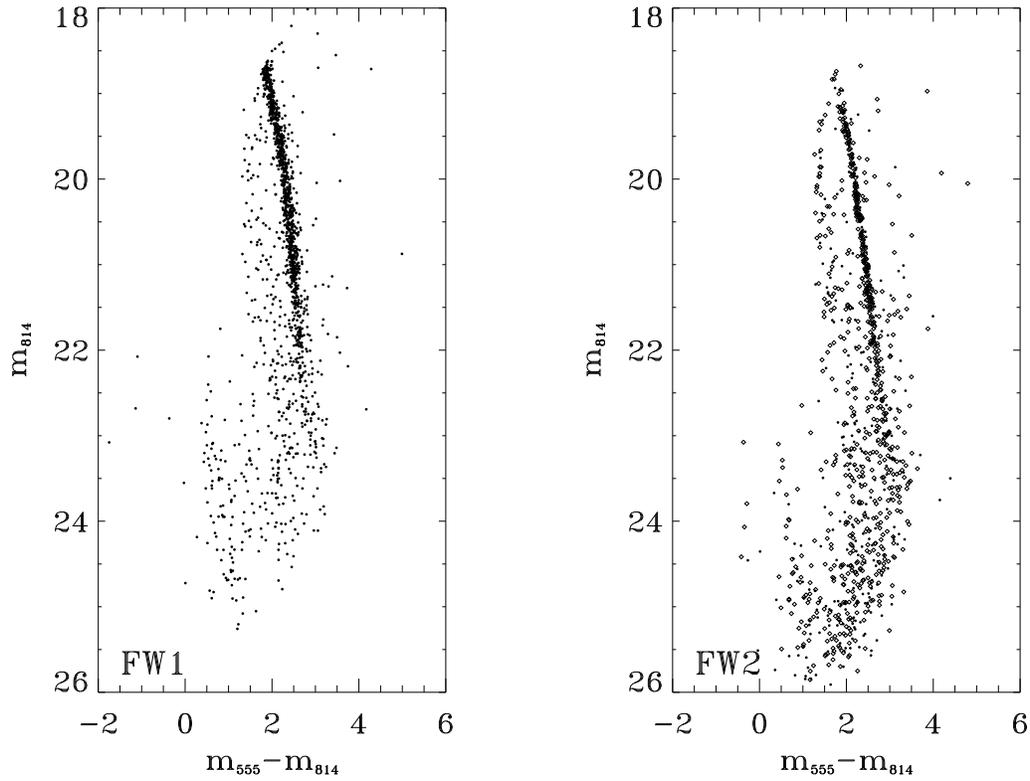}\hfill
\parbox[b]{30mm}{
 \caption[]{
 The $m_{814},m_{555}-m_{814}$ CMD for the stars in the inner field FW\,1
 and in the external field FW\,2}}}}
 \end{figure*}

 The final step in obtaining the LF of M\,4 at these locations in the
 cluster is to subtract, after correction for photometric
 incompleteness, the SKY LF from the LFs measured in FN\,1 and FN\,2. The
 LF derived in this way is shown in Fig.\,10 where the LFs in fields
 FN\,1 (solid line) and FN\,2 (dashed line) are corrected for
 incompleteness and contamination. The more external LF (FN\,2) has been
 shifted vertically to reach the same maximum as the other. Within the
 observational errors, the two LFs are indistinguishable from one
 another.

 \subsection{The WFPC\,2 data}

 HST--WFPC\,2 observations of M\,4 are available in the HST archive,
 from which we have retrieved two sets of F555W, F814W band images
 obtained in 1995 at $\sim 1\minpoint3$\,E (field FW\,1) and $\sim
 5^\prime$\,SE (field FW\,2) of the cluster center (see Fig.\,1).
 Since field FW\,1 contains our NICMOS FN\,1 field, we are able to
 compare the LFs measured at the same location in different bands. In
 addition, field FW\,2 provides information on the variation of the LF
 as a function of the radial distance.

 The raw images were processed using the standard HST pipeline
 calibration. Images of the same field taken through the same filter
 were combined to remove cosmic ray hits and to improve the SNR of the
 data. The total exposure time of the combined images of field FW\,1
 corresponds to $15,000$\,s in the F555W band and $5,500$\,s in the
 F814W band. Exposure times for FW\,2 are $31,500$ \,s and $7,200$\,s
 respectively in F555W and F814W.

 We have measured the fluxes of all the unresolved objects detected in
 these images by using almost the same photometric reduction technique
 employed for the NICMOS data (see Sect.\,2.1), excluding the PC chips.
 In particular, we have run the automated star-detection routine
 daofind on the combined frames to detect objects rising at least
 $5\,\sigma$ above the local average background. Each of the detected
 objects has been visually examined and the spurious identifications,
 due essentially to the diffraction spikes of a number of saturated
 stars, were removed from the output list. In this way a total of
 $1439$ and $1078$ unsaturated objects were detected, respectively in
 fields FW\,1 and FW\,2.

 The photometric routines {\it phot} and {\it allstar} were used to
 measure the fluxes of the objects detected in such a way on the single
 frames. The PSF was modeled by averaging moderately bright stars for
 each frame. The instrumental magnitudes from the different frames were
 weighted by the inverse square of their uncertainties, averaged for
 each filter and converted to the instrumental magnitudes of the
 combined images.

 We calculated the aperture correction for each WF chip in the two
 fields, using a diameter size of 1\arcsec and transformed the final
 instrumental magnitudes into the VEGAMAG photometric system adopting
 the updated zero points listed in the November 1997 edition of the HST
 Data Handbook.

 Fig.\,11 displays the $m_{814},m_{555}-m_{814}$ CMDs for all objects
 of the inner field FW\,1 and of the external field FW\,2. The
 completeness corrections have been estimated by running artificial star
 tests in both bands and in the combined images for each WF chip. The
 procedure is essentially the same as described in Sect.\,2.1. Our
 simulations indicate that in the inner field at $m_{814} \simeq 21$ on
 average about $90\%$ of the artificial stars are recovered. In the
 outer field, the same completeness level is reached at $m_{814} \simeq
 22$ due to the lower crowding.

 \begin{figure*}
\vbox{\hbox
to\textwidth{\psfig{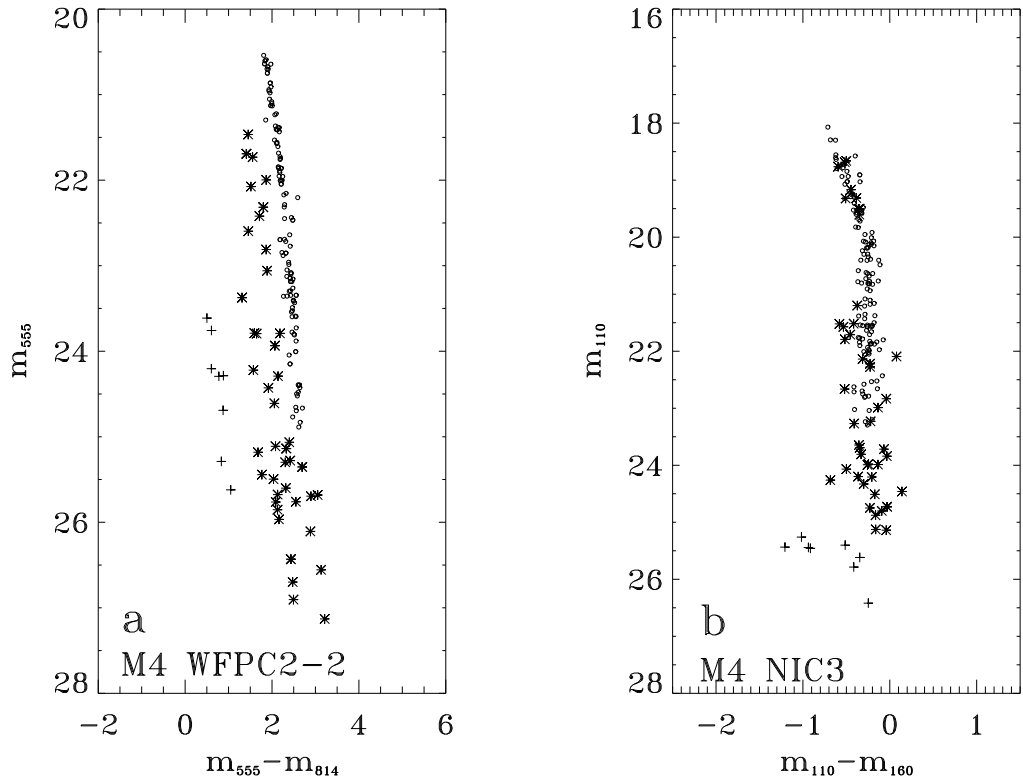}\hfill
\parbox[b]{30mm}{
 \caption[]{{\bf a}~WFPC2 CMD of the 193 objects in common with the
 NICMOS FN\,1 field. {\bf b}~NIC3 CMD of the same objects. Crosses
 represent white dwarfs, asterisks field stars, and circle
 cluster members}}}}
 \end{figure*}

 Fig.\,12 shows the NICMOS and the WFPC\,2 CMDs of the 193 stars common
 to both inner fields. The center of the overlapping region is located
 $\sim 1\minpoint9$\,NE of the cluster center. The crosses represent
 objects already identified as white dwarfs (WD) by Richer et al.
 (1997) on the basis of the agreement between observational and
 theoretical loci of a $0.5$\Msolar Hydrogen-rich WD cooling sequence,
 the asterisks indicate objects which, lying outside the narrow cluster
 MS, probably belong to the foreground whereas the circles mark the
 cluster members.

 As a first result we have the observational confirmation of the
 contamination of the cluster MS in the infrared due to field stars: it
 is impossible to follow the MS in the IR bands all the way to its very
 bottom without a statistical decontamination through the use of an
 external field or excluding the foreground stars using proper motion
 data.

 Secondly, the cross--identification of the stars in both the WFPC\,2
 and NICMOS photometry tables enables us to identify in a globular
 cluster for the first time an infrared WD sequence.  In the F110W band
 the WDs are distributed in a {\it color} sequence instead of a {\it
 luminosity} sequence as in the optical CMD.  This is an important
 observational constraint for the modeling of the atmosphere of cool
 WDs. Bergeron, Saumon, \& Wesemael (1995) have shown that the cool WDs
 with mixed $H/He$ chemical composition can be easily recognized from
 their predicted strong infrared flux deficiency.

 Adopting the $2.5\,\sigma$ clipping criterion (De Marchi \& Paresce
 1995), from the CMDs of Fig.\,11 we have measured the LF of MS stars
 by counting the objects in each $0.5$ mag bin and within $\pm2.5$
 times the $m_{555}-m_{814}$ color standard deviation around the MS
 ridge line. The WFPC\,2 LFs of both observed fields, corrected for
 incompleteness and field star contamination, are shown in Fig.\,13.
 The exposure times of the WFPC2 frames were calculate to reach the
 very end of the MS and, as a consequence, the saturation effect does
 not allow us to evaluate reliable LFs at $m_{814} < 19$ while the
 foreground contamination cuts them at $m_{814} > 22.5$.

 \section{The luminosity and mass functions}

 Fig.\,14 shows the F814W LFs of the two WFPC\,2 fields located at
 $\sim 2\arcmin$ (triangles) and at $\sim 6\arcmin$ (boxes) from the
 cluster center, together with the NICMOS LF (diamonds) converted into
 the F814W band using the Baraffe et al. (1997) M-L relationships
 appropriate for the metallicity of M\,4 ($[M/H] = -1$, i.e. $[Fe/H]
 \simeq -1.3$; Harris 1996) as a function of the absolute $M_{814}$
 magnitude.  Through the theoretical M-L relations, we can associate a
 mass to each of the magnitude bins that appear in the NICMOS LF.
 Then, by also knowing the M-L relation for the F814W band, we can
 easily translate the center of the magnitude bins from the F110W
 magnitude scale to the F814W band and adjust the width of the bins
 themselves to reflect the different slope of the M-L relations in the
 magnitude range covered by each bin.

 Within the error bars, in the overlapping range, the inner NICMOS and
 inner WFPC\,2 LFs show the same trend:  the overall LF is relatively
 flat or slightly decreasing in the range $4 < M_{814} < 8$ and
 decreases steeply beyond that point. The agreement between the two LFs
 measured in the visible (F814W) and in the near IR (F110W), in spite
 of the small size of the observed sample, confirms the robustness of
 our photometric reduction and the reliability of the M--L relations
 used to translate the IR data to the visible (see also De Marchi 1999
 on this issue).

 \begin{figure}
 \vbox{\hbox to\textwidth{\psfig{figure=m4.f13,width=70mm}\hfill}}
 \caption[]{Luminosity functions of the fields FW\,1 (triangles) and
 FW\,2 (squares) after correction for incompleteness and
 field star contamination}
\vspace{5mm}
 \vbox{\hbox to\textwidth{\psfig{figure=m4.f14,width=70mm}\hfill}}
 \caption[]{Logarithm of the WFPC2 $2\arcmin$ (triangles) and
 $6\arcmin$ (boxes) LFs in the F814W band together with the NICMOS LF
 (diamonds) translated to F814W adopting M-L relations of Baraffe et
 al.  (1997). Solid lines represent the fit obtained using a power-law
 with exponent $x = -0.8$ in the inner field.  As regards the outer
 field, the LF is reproduced using a two-segment power-law, with
 exponents $x = 1$ down to $M_{814} \simeq 8.5$ and $x = -0.3$ down to
 the detection limit}
 \end{figure}

 On the other hand, the WFPC\,2 LF measured $6\arcmin$ away from the
 center ($\simeq 1.6 r_{\rm hm}$; Harris 1996) shows a different shape,
 in that it increases with decreasing luminosity up to a peak at
 $M_{814} \simeq 9$ and from there it drops all the way to the MS
 detection limit at $M_{814} \simeq 10.5$. This behavior is typical of
 all the GCs observed so far with the HST at or near the half-mass
 radius (Paresce, De Marchi \& Romaniello 1995; De Marchi \& Paresce
 1995a, 1995b, 1996, 1997; Cool, Piotto, \& King 1996; King et al.
 1997; Pulone, et al. 1998, De Marchi 1999) and is related to an
 extremum in the derivative of the mass-absolute magnitude relation of
 the low mass stars as discussed in Kroupa \& Tout (1997).

 Thus, the obvious difference between the inner and outer LFs should be
 compatible with the dynamical modifications of a GC as a result of
 mass segregation due to energy equipartition through two-body
 relaxation (Spitzer 1987). Indeed, it is expected that more massive
 stars give up part of their kinetic energy to lighter stars during
 close encounters and sink towards the cluster center.  This effect
 would explain qualitatively why massive stars are relatively more
 numerous at $2\arcmin$ from the center than they are near the
 half-mass radius. We address this issue in more detail in Sect.\,4.

 Under the assumption that the MF of the stellar population at
 $2\arcmin$ from the cluster center is represented by an exponential
 distribution of the form:
 \begin{equation}
 dN \propto m^{-x} d\log m
 \end{equation}
 and using the M-L relation of Baraffe et al., we can reproduce rather
 accurately the inner LF in Fig.\,14 with an exponent $x = -0.8 \pm
 0.4$ (solid line). The outer LF, however, cannot be fit by any single
 exponent power-law distribution, regardless of the adopted slope, as no
 such function can reproduce the peak observed at $M_{814} \simeq 9$.
 A reasonable fit can be obtained using a mass function which rises as a
 power-law with $x = 1$ down to $\sim 0.25 M_\odot$ ($M_{814} = 8.5$)
 and then drops all the way to the MS detection limit with an exponent
 $x = -0.3$. Unfortunately the lack of data at the bright end of the
 WFPC2 LF, due to saturation, does not allow us to constrain the MF
 slope in the mass interval which is most sensitive to dynamical
 effects, although the comparison between the inner and outer mass
 functions at lower masses already suggests strong mass segregation.

 \section{The dynamical state of M\,4}

 Addressing the question as to whether the difference between the inner
 and outer MFs described above can be quantitatively accounted for by
 the two-body relaxation mechanism requires a simulation of the
 dynamical structure of the cluster. We have used multi-mass
 King-Michie models, constructed with an approach nearly identical to
 that of Gunn \& Griffin (1979), as described in Meylan (1987, 1988).
 Each model is characterized by an IMF in the form of an exponential as
 in Eq.\,(3), with an exponent $x$ which is not necessarily constant
 (i.e. $x$ is allowed to vary with mass; see details below), and by
 three structural parameters describing respectively the scale radius
 ($ r_{\rm c}$), the scale velocity ($v_{\rm s}$) and the central value
 of the dimensionless gravitational potential $W_{\rm o}$.  We have
 assumed complete isotropy in the velocity distribution.

 Because King-Michie modeling provides a ``snapshot'' of the current
 dynamical status of the cluster rather than its evolution, it is
 useful to define the mass distribution of cluster stars at present
 (global MF; GMF) as the MF that the cluster would have simply as a
 result of stellar evolution (i.e. ignoring any local modifications
 induced by internal dynamics and/or the interaction with the Galactic
 tidal field). Clearly, the IMF and GMF of MS (unevolved) stars is the
 same.  For practical purposes, the GMF has been divided into sixteen
 different mass classes, covering MS stars, white dwarfs, and heavy
 remnants. All stars lighter than $0.8 M_{\odot}$ have been considered
 still on their MS, while heavier stars with initial masses in the
 range $[8\mbox{--}100]$\,M$_{\odot}$ have been assigned a final (i.e.
 post-evolution) mass of $1.4$\,M$_{\odot}$. White dwarfs have been
 divided into three classes, according to their initial masses,
 following the prescriptions of Weidemann (1987) and Bragaglia,
 Renzini, \& Bergeron (1995). Although the analytical form we adopted
 for the IMF would allow the exponent $x$ of the GMF to vary, we have
 initially restricted our investigation to the single exponent case.

 From the parameter space defined in this way, we have extracted only
 those models that simultaneously fit both the observed surface
 brightness (SBP) and velocity dispersion (VDP) profiles of the
 cluster. The fit to the SBP and VDP, however, can only constrain
 $r_{\rm c}$, $v_{\rm s}$, and $W_{\rm o}$, while still allowing the
 IMF to take on a variety of shapes. It is precisely to attempt to
 break this degeneracy that we impose the condition that the model GMF
 agree with the observed LFs.  We note here that the local MF at any
 given place inside the cluster originates from the GMF and stems
 directly from imposing the condition that stars of different masses be
 in thermal equilibrium with each other.

 In our simulations, we have used the VDP determined by Peterson, Rees,
 \& Cudworth (1995).  Concerning the SBP, we have compared the profile
 measured by Kron, Hewitt, \& Wasserman (1984) with that of Trager,
 King, \& Djorgorvski (1995) and found some discrepancies. We have,
 therefore, adopted an average SBP, defined as the parameterized
 King-type profile (King 1962) that corresponds to the {\it canonical}
 values of core radius $r_{\rm c} = 50$\arcsec and concentration $c =
 1.6$ (Harris 1996; note, however, that values as large as $r_{\rm c}
 \simeq 90$\arcsec and $c \simeq 2$ are found in the literature, as
 discussed below).  The {\it canonical} SBP defined in this way gives a
 reasonable fit to both measurements, and we have used the least
 squares between the canonical profile and the observations of Kron et
 al. (1984) and Trager, King, \& Djorgorvski (1995) as an estimate of
 the errors.

 Although several models are able to reproduce the observed radial
 profiles (SBP and VDP) reasonably well, no choice of the exponent $x$
 gives local MFs that, converted into LFs using the M-L relation
 described above, agree with the data shown in Fig.\,14. In fact, our
 simulations show that an IMF like the one given in Eq.\,(3) cannot
 simultaneously fit the LFs at $2\arcmin$ and $6\arcmin$ radial
 distance, regardless of the value of the exponent $x$, although each LF
 can individually be approximated with a power-law IMF with index $x
 \simeq -0.8$ for the LFs at $2\arcmin$ radius and $x \simeq 0.4$ at
 $6\arcmin$. (In the latter case, however, the fit is not good at
 magnitudes fainter than $M_{814} \simeq 9$). As an example, in
 Fig.\,15 we show the theoretical LFs (solid lines) corresponding to
 the model that best fits the observations at $6\arcmin$: at bright
 magnitudes (i.e. for masses $> 0.3$\,\Msolar) the model LF systematically
 deviates from the observational data at $2\arcmin$ by far more than
 the experimental errors.

 \begin{figure}
 \vbox{\hbox to\textwidth{\psfig{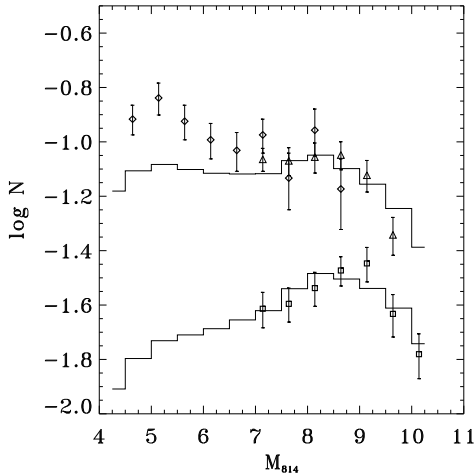}\hfill}}
 \caption[]{The WFPC2 $2$\arcmin (triangles) and $6$\arcmin(squares)
 LFs, the NICMOS LF (diamonds) in the F814W band, compared with the
 theoretical LFs at the same distances, obtained by the mass
 stratification of the dynamical model described in the text. The slope
 of the main sequence mass segment of the GMF is $x = 0.4$.  The core
 radius and the concentration of the parameterized SBP are $r_{rm c} =
 50$\arcsec and $c = 1.6$}
 \end{figure}

 A single-exponent power-law distribution, however, is only a rough
 approximation to the IMF, particularly at the low-mass end (see Scalo
 1998). And indeed, the deepest LFs today available for NGC\,6397
 (Paresce, De Marchi, \& Romaniello 1995; King et al. 1998) and
 NGC\,6656 (De Marchi \& Paresce 1997) suggest that the MF could
 flatten out below $\sim 0.25$\,\Msolar and possibly drop (in the
 logarithmic plane). We have, therefore, run some simulations with an
 IMF that follows an exponential law with two values of the exponent:
 $x$ is positive for masses $m > 0.25$\,\Msolar and zero below that
 mass. Not even in this case, however, does the predicted shape of the
 MF at $\sim 2\arcmin$ and $\sim 6\arcmin$ from the center match the
 observations simultaneously, regardless of the value of $x$ for masses
 $m < 0.25$\,\Msolar.

 Although we could explore other possible shapes of the IMF, it is
 unlikely that its form can largely deviate from some sort of a slowly
 varying exponential. Thus, if we judge that King-Michie modeling is a
 viable way to describe the dynamical state of M\,4, we have to
 conclude that some of the constraints that we are forcing into the
 models are not correct, i.e. that the SBP or the VDP (or both) are not
 compatible with the LFs that we measured and which seem rather robust
 (see Sect.\,3).

 We have, therefore, investigated the origin of the discrepancy between
 the observed LFs and the local MFs produced by the models starting
 from plots similar to that shown in Fig.\,15. We have noticed that, if
 we begin by fixing the exponent (exponents) of the IMF in such a way
 that the local MF agrees with our measurements at $6\arcmin$ from the
 center, we are tempted to conclude that the scale radius of our models
 is too small.  Fixing the exponent in that way is a reasonable
 assumption, since both theoretical and semiempirical arguments suggest
 that near the cluster half-mass radius ($\sim$ half-light $\sim
 3.7$\arcmin; Harris 1996), the local MF should only marginally differ
 from the IMF (Richer et al. 1991; Vesperini \& Heggie 1997).

 In fact, in order for the model LF to reproduce the LFs measured at
 $2\arcmin$, the former has to decrease monotonically with luminosity,
 and this is precisely what has been shown to take place in the
 innermost cluster regions (see Paresce, De Marchi, \& Jedrzejewski
 1995; King, Sosin, \& Cool 1995; De Marchi \& Paresce 1996). In other
 words, we would need a scale radius ($r_{\rm c}$) sufficiently large
 so as to place the region at $r \simeq 2$\arcmin \,still close enough
 to the center that the LF of its stellar population is compatible with
 an inverted function. The canonical value of $r_c = 50$\arcsec is not
 large enough for this to happen.

 \begin{figure}
 \vbox{\hbox to\textwidth{\psfig{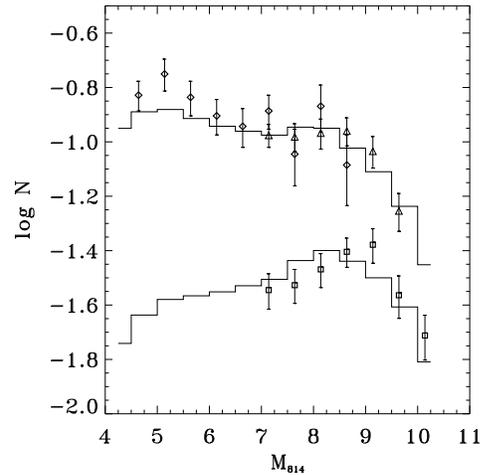}\hfill}}
 \caption[]{
 The same as Fig.\,15 when adopting as GMF a power-law with
 exponents $x = 0.2$, in the range $0.8\mbox{--}0.25$\,\Msolar,
 and $x = -0.4$ for masses $< 0.25$\,\Msolar}
 \end{figure}

 \begin{figure}
 \vbox{\hbox to\textwidth{\psfig{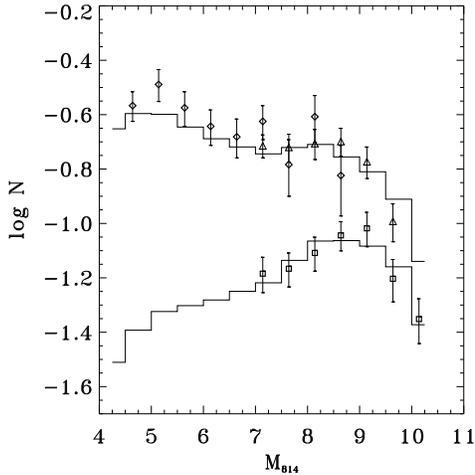}\hfill}}
 \caption{The same as Fig.\,15 when the radial LFs are fitting a SBP
 with $r_{\rm c} = 92$\arcsec and $c = 1.9$.  The adopted GMF has main
 sequence slopes $x = 1$, in the range $0.8\mbox{--}0.25$\,\Msolar, and
 $x = -0.4$ for masses $< 0.25$\,\Msolar}
 \end{figure}

 Through repeated trials, we have found that by forcing the core radius
 $r_{\rm c}$ to take on the value of $\sim 90$\arcsec (a figure still
 compatible with the measurements of Peterson \& King (1975) suggesting
 $r_{\rm c} \simeq 93$\arcsec from the fit to their SBP) both sets of
 LFs can be simultaneously fit by an exponential IMF with index $x
 \simeq 0.2$ in the range $0.8\mbox{--}0.25$\,\Msolar and which drops
 with $x \simeq -0.4$ at lower masses (Fig.\,16). The values of the
 concentration ratio ($c \simeq 1.5$), the total cluster mass($\simeq
 8.8 \times 10^4$\,\Msolar), and the index of the IMF above
 $0.8$\,\Msolar ($x \simeq 1.4$) are also reasonable. In the mass range
 $0.2 - 0.6$\,\Msolar, the slope of the IMF suggested by this model is
 shallower than the $x \simeq 1$, value proposed by De Marchi \&
 Paresce (1997) for NGC\,6397 and NGC\,6656. Nevertheless, it would
 still be possible to find a model (Fig.\,17) that uses the IMF of De
 Marchi \& Paresce (1997) and simultaneously fits the LFs observed at
 $\sim 2\arcmin$ and $\sim 6\arcmin$, and this would require the core
 radius to grow to $\sim 90$\arcsec and the concentration parameter to
 take on the value of $c \simeq 1.9$ (both values still compatible with
 the data of Peterson \& King (1975) and Kron et al. (1984),
 respectively). The difference between the IMFs used in Figs.\,16 and
 17 and between the set of parameters defining our models can be taken
 as an indication of the uncertainties affecting our approach, and can
 be traced back to the indetermination accompanying the SBP and the
 ensuing value of $r_{\rm c}$. It also shows how difficult it is to
 infer the shape of the GMF starting from the observations when the
 knowledge of the structural parameters of the cluster is so
 uncertain.

 We would also like to point out that the results given above and shown
 in Figs.\,16 and 17 should anyhow be taken with care: they stem from
 the assumption that King-Michie modeling with the limitations that we
 applied to it is adequate to describe the dynamical state of M\,4, and
 therefore the good fits that we obtain do not necessarily guarantee
 that these models reflect the physical state of the cluster.  And
 indeed, from the kinematics of its orbit (Dauphole et al. 1996) one
 would argue that M\,4 should have undergone extensive tidal stripping
 and shocks since it penetrates very deeply into the Galactic bulge and
 its orbital plane is one of the closest to the Galactic disk. These
 effects might have strongly modified the local mass distribution
 (particularly in the outer regions) and the two-body relaxation
 mechanism that we adopted could no longer apply.  We are grateful to
 Pavel Kroupa, the referee, for having pointed out that, among the
 possible problems with the assumptions underlying our modeling, there
 is the assumed complete isotropy in the velocity distribution of the
 cluster members.  Stellar-dynamical modeling of isolated globular
 clusters shows that the velocity dispersion becomes anisotropic as a
 function of time (see Fig.\,4e in Spurzem \& Aarseth 1996).  Models
 that include a tidal field, however, suggest that the anisotropy is
 reduced because of rapid loss of stars having radial orbits
 (Takahashi, Lee, \& Inagaki 1997).  Nevertheless, it is not clear if
 the tidal field is realistically taken account of in the Takahashi et
 al. (1997) Fokker-Planck models, since in the above mentioned Fig.\,4e
 of Spurzem \& Aarseth (1996) it can be seen that the anisotropy would
 also be pronounced well inside the tidal radius. Moreover, GCs may be
 born with a significant anisotropic velocity dispersion (Aarseth, Lin,
 \& Papaloizu 1988).

 With this caveat in mind, we can quantitatively conclude that the LFs
 observed at $\sim 2\arcmin$ and $\sim 6\arcmin$ away from the center
 of M\,4 can be traced back to the same GMF, which appears to have been
 locally modified as a result of mass segregation due to two body
 relaxation. The underlying IMF is most likely represented by a slowly
 rising exponential ($x \simeq 0.2\mbox{--}1$) that flattens out and
 possibly drops below $\sim 0.25$\,\Msolar. Any attempt to better
 define the shape of the IMF and to characterize more thoroughly the
 relaxation mechanism would require a more precise measurement of the
 surface brightness (or, better, radial density) profile and the
 observation of the LF of MS stars at four or five locations inside the
 cluster, so as to simultaneously constrain both the core radius (for a
 given mass class) and the mass spectrum of all luminous objects as a
 function of the distance from the center. This project can be
 efficiently and effectively carried out with a wide field imager at a
 10-m class telescope like the VLT.

 \section{Summary}
 
 Our main conclusions can be summarized as follows:
 
 \begin {enumerate}

 \item We have measured the visible and infrared LFs of M\,4 at
 $2$\arcmin and $6$\arcmin \,from the cluster center down to $I \simeq
 10$ with photometric completeness always better than 50\%.

 \item The LFs measured at $2\arcmin$ with the two instruments agree
 very well with each other and show a slow decrease with luminosity.

 \item The LF measured with the WFPC2 at $6$\arcmin\,radial distance
 shows an increase with decreasing luminosity up to a peak at $I \simeq
 9$ followed by a drop all the way to the detection limit.

 \item The difference between the two sets of LFs is qualitatively
 compatible with the effects of mass segregation.  It can, however, be
 quantitatively reproduced using a King-Michie model only when adopting
 cluster structural parameters which, although compatible with those
 available in the literature, deviate from their {\it canonical}
 value.  The IMF required for the model to fit has an exponential slope
 of $\sim 0.2\mbox{--}1.0$ in logarithmic mass units between $0.2$ and
 $0.8$ \Msolar.

 \item Our observations require that the IMF determined in this way
 flatten out and possibly drop below $0.2$ \Msolar.

 \item For the first time an infrared WD sequence has been
 unambiguously identified in a globular cluster.  The WD locus appears
 to be a {\it color} sequence instead of a {\it luminosity} sequence as
 in the optical.

 \end {enumerate}

 \begin{acknowledgements}

 We would like to thank Pavel Kroupa, the referee of this work, for his
 critical comments that have greatly improved our paper.  Many thanks
 are due to Gilles Chabrier and Isabelle Baraffe for contributing their
 theoretical M-L relations and for helpful comments and suggestions. We
 are indebted to George Meylan for many helpful discussions on the
 dynamical modeling of globular clusters and to Paolo Montegriffo and
 Donata Guarnieri for their expert assistance in setting up the fortran
 code.
 
 \end{acknowledgements}

 \end{document}